\documentclass[fleqn,twoside]{article}
\usepackage[headings]{espcrc2}
\usepackage{fancyvrb}
\readRCS
$Id: espcrc2.tex,v 1.2 2004/02/24 11:22:11 spepping Exp $
\usepackage{graphicx, balance}
\usepackage[figuresright]{rotating}
\usepackage{subfigure}

\usepackage{authblk}
\newcommand{\AmS}{{\protect\the\textfont2
  A\kern-.1667em\lower.5ex\hbox{M}\kern-.125emS}}

\title{\textbf{Design of Novel Architectures and FPGA Implementation of 2D Gaussian Surround Function}}

\author{M. C Hanumantharaju and M. T Gopalakrishna \\{Department of Information Science and Engineering, Dayananda Sagar College of Engineering, Bangalore 560 078, India, Contact: mchanumantharaju@gmail.com, gopalmtm@gmail.com \\}
}
\runtitle{International Journal of Information Processing}
\runauthor{M. C Hanumantharaju, et al.,}

\begin{document}
\begin{abstract}
A new design and novel architecture suitable for FPGA/ASIC implementation of a 2D Gaussian surround function for image processing application is presented in this paper. The proposed scheme results in enormous savings of memory normally required for 2D Gaussian function implementation. In the present work, the Gaussian symmetric characteristics which quickly falls off toward plus/minus infinity has been used in order to save the memory. The 2D Gaussian function implementation is presented for use in applications such as image enhancement, smoothing, edge detection and filtering etc. The FPGA implementation of the proposed 2D Gaussian function is capable of processing (blurring, smoothing, and convolution) high resolution color pictures of size up to $1600\times1200$ pixels at the real time video rate of 30 frames/sec. The Gaussian design exploited here has been used in the core part of retinex based color image enhancement. Therefore, the design presented produces Gaussian output with three different scales, namely, 16, 64 and 128. The design was coded in Verilog, a popular hardware design language used in industries, conforming to RTL coding guidelines and fits onto a single chip with a gate count utilization of 89,213 gates. Experimental results presented confirms that the proposed method offers a new approach for development of large sized Gaussian pyramid while reducing the on-chip memory utilization.   \\\\

{\bf Keywords :} Gaussian Surround Function, Hardware Architecture, FPGA, Verilog.
\end{abstract}

\maketitle

\section{Introduction}

With the widespread use of technologies like digital television, internet streaming video and DVD video, Gaussian function has become an inevitable component of image/video processing \cite{tinku} and pattern recognition. Hardware realization of 2D Gaussian surround function for image processing applications demands huge on-chip memory requirement, with massive computations. Further, these functions might not fit on a single FPGA device. This is due to the reason that Gaussian function has an exponential distribution with maximum entropy and implementation such functions using software schemes are complex from computation point of view. In addition, Gaussian function is a non-causal, which implies that function is symmetric about the origin in time domain. Gaussian filter implementation with smaller kernel size in order to blur an image have been reported by number of researchers \cite{ngo}. However, design of Gaussian function for large kernel size requires enormous amount of resources on FPGA with incredible raise in the total equivalent gate count.   

The new approach for 2D Gaussian function design provides a technical solution appropriate for the broad range of applications, from image pre-processing to pattern recognition. It simplifies the computational complexity with considerable saving in the memory requirement. The work proposed here utilizes the symmetry property of Gaussian surround function in order to save hardware resource, particularly memory requirement on FPGA. The technical design presented in this work is highly focused on providing huge throughput to process images as well as motion pictures. The Gaussian surround function is designed to process high quality images with picture size exceeding $1600\times1200$ using reduced memory and high speed computation. 

The rest of the paper is organized as follows: Section 2 presents the review of related work. Section 3 describes the proposed 2D Gaussian surround function design. Section 4 gives brief details of architecture development suitable for FPGA/ASIC implementation. Section 5 provides experimental results and discussions. Finally conclusion arrived at is presented in Section 6. 

\section{Related Work}
Image and Video Processing has been a very active field of research and development for over 20 years and many different systems and algorithms for image enhancement, restoration, filtering and smoothing have been proposed and developed. The core part in all these image processing techniques is the Gaussian function. Although a lot of research work has progressed in the development of Gaussian design for image processing application using software and hardware schemes, very little work seems to have been carried out in large sized Gaussian function design. In addition, there are no full-fledged implementations reported for 2D Gaussian surround function using FPGA or ASIC that demands the development of novel algorithms for high speed processing and the best possible memory utilization. The market demands for 2D Gaussian function design are very high. The design of 2D Gaussian function and architecture development \cite{donald} is not only challenging but also intellectually stimulating. These are the primary reasons why this work was undertaken by the present researcher.

Jobson et al. \cite{journ01} described the properties of center or surround retinex. In this work authors have claimed that the design of a Gaussian surround function with a space constant of 80 pixels is a reasonable compromise between dynamic range and rendition. Gaussian function design and its application to retinex based image enhancent with DSP implementation has been proposed by Glenn et al. \cite{dsp_gauss}. DSPs can be employed for enhancement of images which provides some improvement compared to general purpose computers. DSPs can be programmed in different languages, such as assembly codes and C language. The Hardware knowledge is required for programming DSPs, However, it is much easier
for designers to learn DSP programming compared with the other design choices. However, image enhancement algorithms involving Gaussian function developed for DSP implementation may not be parallelized without using multiple DSPs. The DSPs are well suited for implementation of floating point systems, while for ASICs and FPGAs, floating point
operations are difficult to implement. The enhancement of 25-30 frames per second of large size video frames with $1024\times1024$ pixel resolution is still not possible with DSPs. The image enhancement technique requires massive parallel processing capability in order to support real time enhancement of large video stream.

Hiroshi et al. \cite{hiroshi} proposed a real time retinex video image enhancement algorithm and its FPGA implementation. Although the retinex algorithm \cite{tao} has been used for video enhancement the Gaussian pyramid designed in this work is of size $3\times3$. The authors have claimed that the architectures developed in this scheme are efficient and can handle color picture of size $1900\times1200$ pixels at the real time video rate of 60 frames per sec. In this scheme, the computational cost of the algorithm depends on the number of processing layers while the maximum layers and iterations used are 5 and 30 respectively. The authors have not justified how high throughput has been achieved in spite of time consuming iterations to the tune of 30. Further, the algorithm uses HSV color space for the enhancement process. This leads to an additional computational cost with the maximum conversion error occurring in the conversion process from RGB to HSV.
 
Tsung et al. \cite{tsung} proposed algorithm and architecture design of human machine interaction in foreground detection of dynamic scene. The Gaussian function in this work is computed based on fixed point data. This method adapts 27 bits for the variance computation of which 7 bits represents integer portion and 20 bits are used for fraction part. Three look-up tables are used to index the exponential and divide value. However, the Gaussian density function developed here uses large memory and is not efficient from computation point of view. 

Design of novel algorithm and architecture for Gaussian based color image enhancement for real time applications have been proposed by Hanumantharaju et al. \cite{conf01}. The $5\times5$ Gaussian kernel exploited in this work performs the image enhancement operation efficiently. Although the medium sized kernel employed in this work operates on picture size of $1600\times1200$ pixels based on window operation, resulting images are not satisfactory due to presence of halo artifacts in the reconstructed images. Design of large sized Gaussian function in place of $5\times5$ kernel improves visual quality of the enhanced image. Further, this approach consumes enormous amount of FPGA resources. The proposed approach of Gaussian surround function design with reduced on-chip memory utilization is an ideal choice compared to kernel based implementations. In addition, design of large size Gaussian surround function and development of architecture suitable for FPGA/ASIC implementation is the first-of-its-kind in the literature. 

\section{Design of 2D Gaussian Surround Function}

The analog Gaussian function in 1D, 2D and ND is expressed as follows:

\begin{equation} 
G_{1D}(x) = \frac{1}{\sqrt{2\Pi}\sigma}e^{-\frac{x^2}{2\sigma^2}}
\end{equation}

\begin{equation} 
G_{2D}(x, y) = \frac{1}{\sqrt{2\Pi}\sigma}e^{-\frac{x^2+y^2}{2\sigma^2}}
\end{equation}

\begin{equation} 
G_{ND}(\vec{x}) = \frac{1}{\sqrt{2\Pi}\sigma}e^{-\frac{\left|\vec{x}\right|^2}{2\sigma^2}}
\end{equation}

$\sigma$ indicates the width of the Gaussian function. According to statistics, if Gaussian probability density function is considered, than $\sigma$ is referred to as standard deviation and the square of it ($\sigma^2$) is called variance.

The discrete version of 2D Gaussian surround function is given by Eqn. (4).

\begin{equation}
\label{eq116}
G_n(x, y) = K_n \times e^{-\frac{(x^{2} + y^{2})}{2\sigma^{2}}}
\end{equation}

and $K_n$ is given by the Eqn. (5)

\begin{equation}
K_n = \frac{1}{\sum_{i=1}^{M}\sum_{j=1}^{N}{e^{-\frac{x^{2} + y^{2}}{2\sigma^{2}}}}}
\end{equation}

where x and y signify the spatial coordinates, $M \times N$ represents the kernel size, n is preferred as 1, 2 and 3 since the Gaussian function is designed with three scales, namely, 16, 64, and 128.

The spatial co-ordinate 'x' shown in Eqn. (4) are derived from the vector x which is presented in Figure 1. The spatial co-ordinate 'y' is obtained similar to that of co-ordinate 'x'. It may be observed from the Figures 1 and 2 that the co-ordinates of Gaussian surround function exhibit symmetry property around the origin. Therefore, the spatial co-ordinate for implementation of Gaussian surround functions can be easily stored in the memory. The graphical plots of Gaussian surround function with the scales of 16, 64 and 128 are shown in Figures 3(a), (b) and (c), respectively.   

\begin{figure}
\centering
{
\includegraphics[height=2in,width=2.5in]{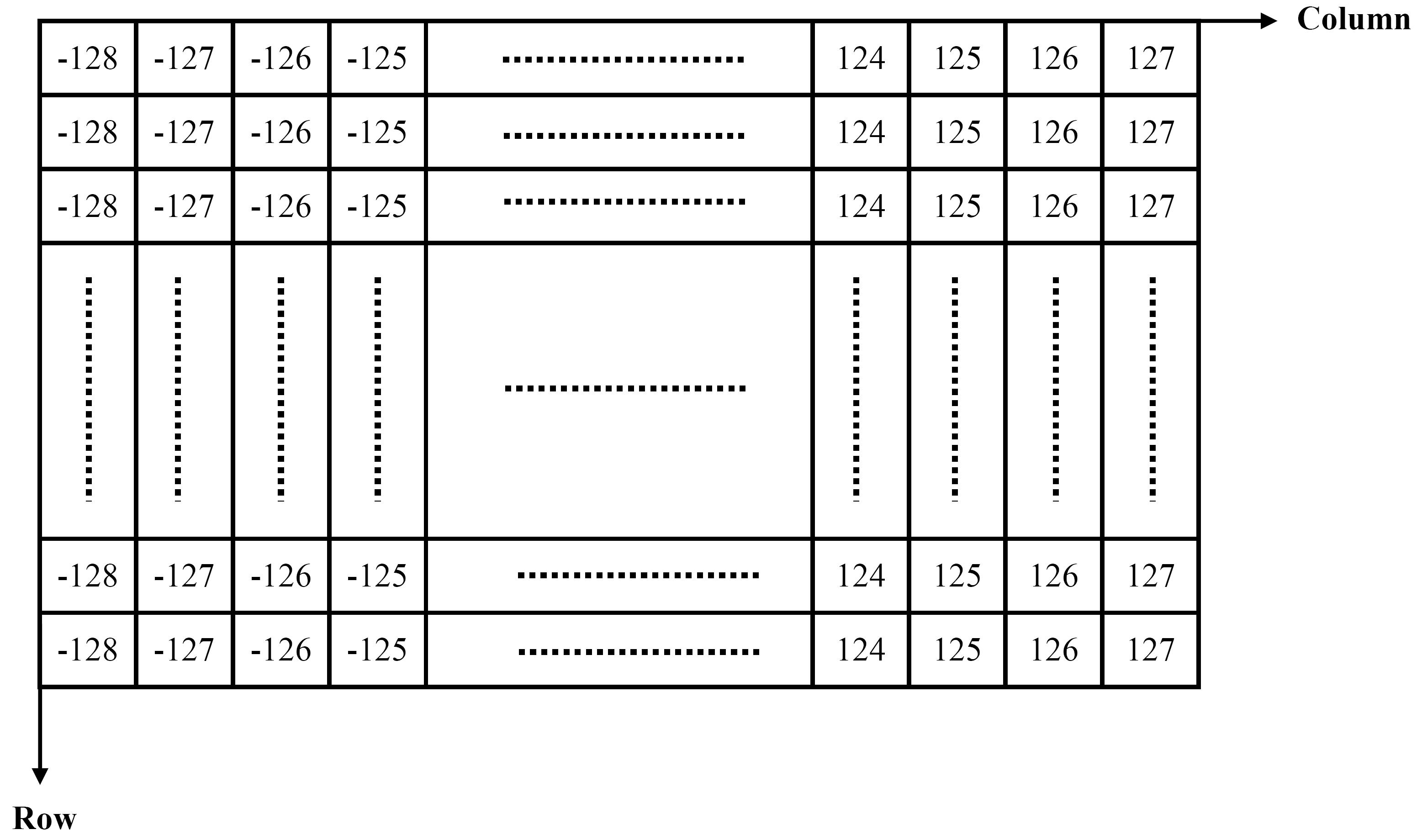}
\label{fig:subfig3}
}
\caption{Domain Specified by 'x' Vector Transformed into Array 'x'}
\end{figure}

\begin{figure}
\centering
{
\includegraphics[height=2in,width=2.5in]{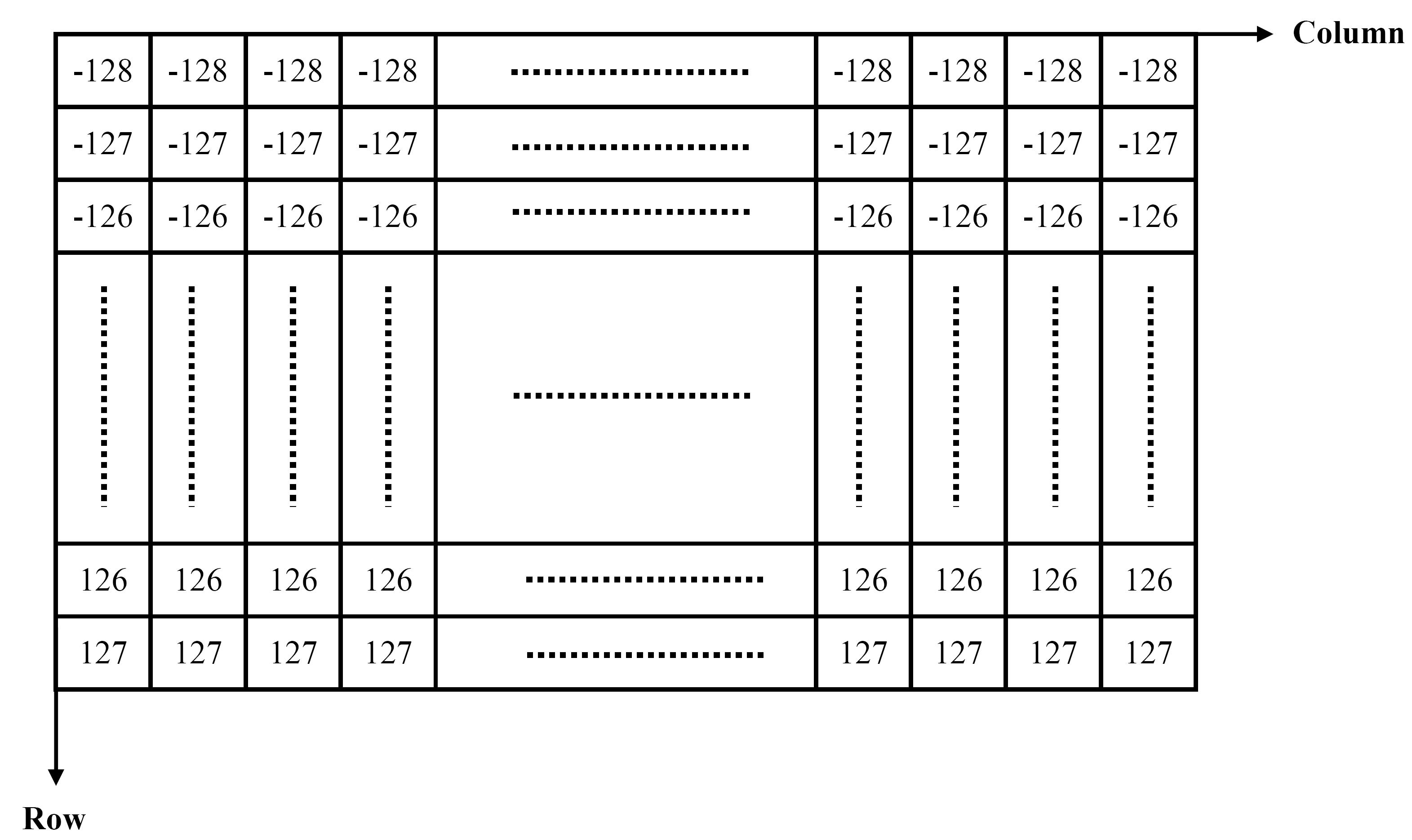}
\label{fig:subfig3}
}
\caption{Domain Specified by 'y' Vector Transformed into Array 'y'}
\end{figure}

% Graphical Plot of Gaussian Function

\begin{figure}
\centering
\subfigure[Gaussian Function with Small Scale]{
\includegraphics[height=0.8in,width=1in]{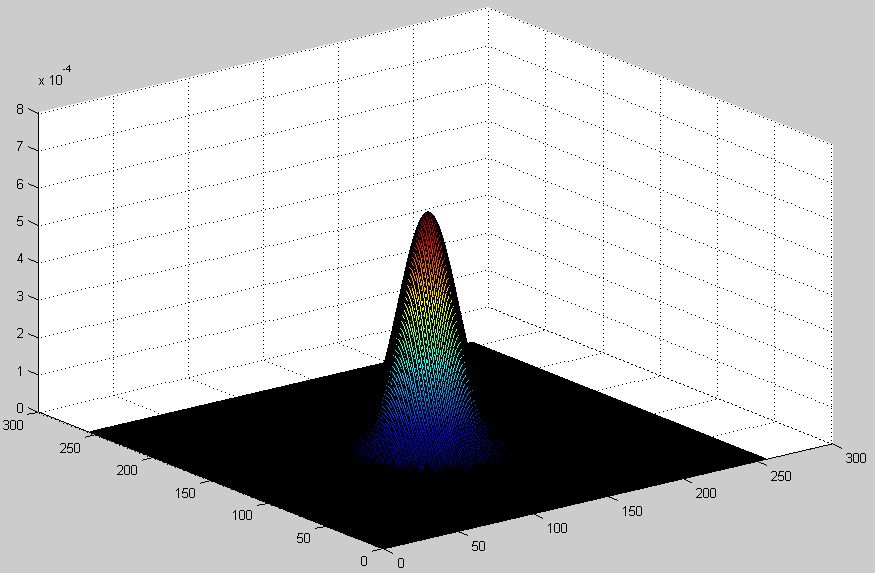}
}
\subfigure[Gaussian Function with Medium Scale]{
\includegraphics[height=0.8in,width=1in]{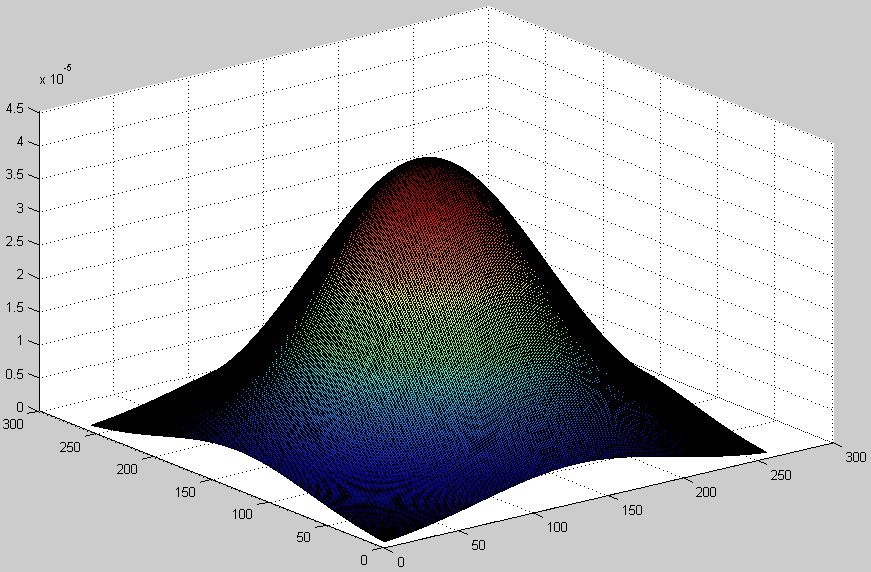}
}
\subfigure[Gaussian Function with Large Scale]{
\includegraphics[height=0.8in,width=1in]{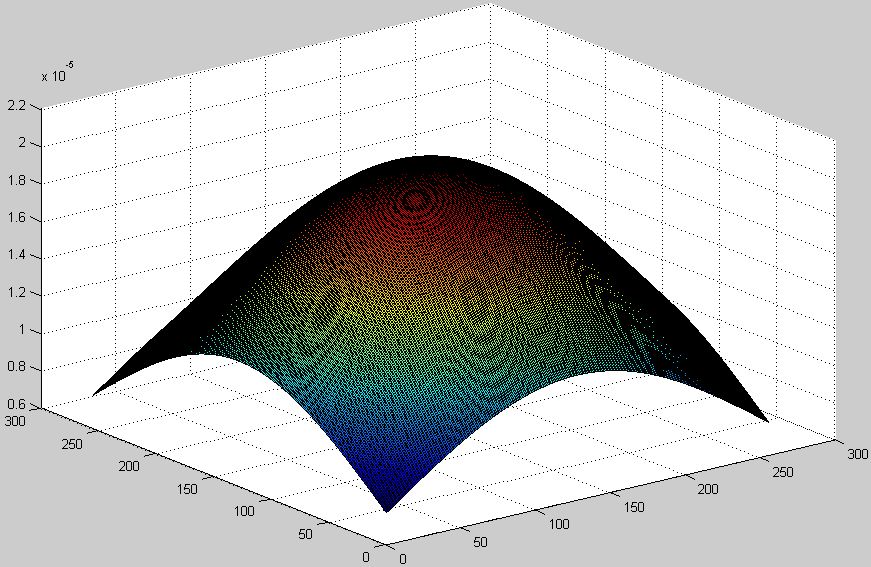}
}
\caption{2D Gaussian Surround Functions with Different Scales}
\label{fig}
\end{figure}

\section{Architectures of 2D Gaussian Surround Function}
The top level architecture of Gaussian surround function comprises counter, dual port ROM, multiplier, scaler and exponential. The overall architecture of 2D Gaussian surround function is presented in Figure 4. The signal description of complete design is shown in Table 1. The top design is called as "gauss2D" its block diagram consists signals "clk", "reset\_n", "start" and "gout [7:0]". The signal "gout [7:0]" is the Gaussian output which is produced at every rising edge of "clk" signal. The signal "reset\_n" is the global reset signal which is used to reset the system during power on conditions. The hardware realization of "gauss2D" functional modules in this work is based on adapting enormous amount of pipeline stages mainly to reduce the computation time. In addition to pipelining technique, parallel processing also has been employed to accelerate the Gaussian function. The Table 2, provides the memory specification of the dual port ROM for the Gaussian surround function of size $256\times256$. The ROM address ranges between 00000000 (corresponds to decimal 0) and 11111111 (corresponds to decimal 255). The contents of the ROM for each location are specified in the second column of Table 2. As is evident from the Table 2, the ROM stores the first row of the spatial co-ordinate 'x' presented in Figure 1. It is possible to produce the array 'x', from the ROM contents since the elements are repeated in the second and subsequent rows of 'x' co-ordinate. Similarly, it is easy to construct the array 'y' presented in Figure 2 by incrementing the address (addr2 [7:0]) of ROM every 256 clock cycle.       

\begin{table*}[ht]
\begin{center}
\caption{Signal Description for the Gaussian Surround Function Design}
\begin{tabular}{|l|l|}
\hline $\textbf{Signals}$ &  $\textbf{Description}$     \\
\hline
clk & This is the global clock signal  \\
\hline
reset\_n & Active low system reset  \\
\hline
start & Asserted to initiate Gaussian function  \\
\hline
gout [7:0] & Gaussian Output  \\
\hline
enable & Asserted to initiate Counting\\
\hline
cnt\_out1 [7:0] and cnt\_out2 [7:0]& Counter Outputs\\
\hline
addr1 [7:0] and addr2 [7:0] & Address for Data selection in ROM \\
\hline
dout1 [7:0] and dout2 [7:0] & Output of Dual port ROM  \\
\hline
n1 [7:0] and n2 [7:0] & Multiplier inputs of 8-bit  \\
\hline
result [15:0] & Multiplier Output of 16-bit  \\
\hline
\end{tabular}
\label{table301}
\end{center}
\end{table*}

\begin{table}[ht]
\begin{center}
\caption{Memory Specification for 2D Gaussian Surround Function}
\begin{tabular}{|l|l|}
\hline $\textbf{Address}$ &  $\textbf{Data}$     \\
\hline
00000000 & 10000000 (-128)  \\
\hline
00000001 & 10000001 (-127)  \\
\hline
00000010 & 10000010 (-126)  \\
\hline
00000011 & 10000011 (-125)  \\
\hline
& \\
\hline
& \\
\hline
& \\
\hline
11111110 & 01111110 (126)  \\
\hline
11111111 & 01111111 (127)  \\
\hline
\end{tabular}
\label{table301}
\end{center}
\end{table}

\begin{figure}
\centering
{
\includegraphics[scale=0.15]{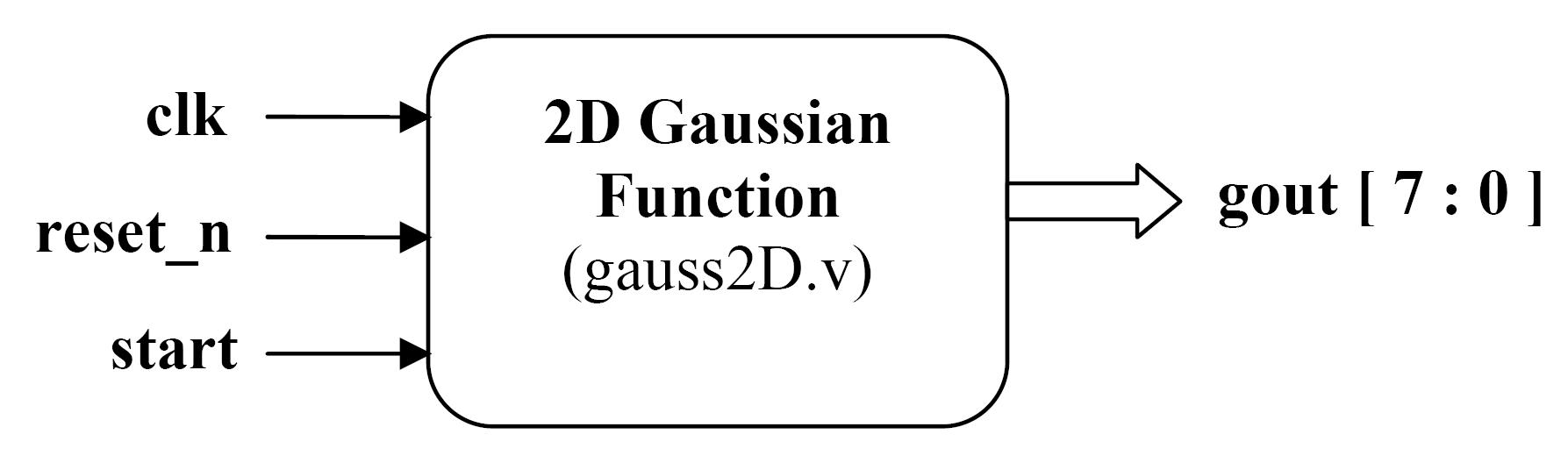}
\label{fig:subfig3}
}
\caption{Top Architecture of 2D Gaussian Surround Function}
\end{figure}

The counters used in the present work is of width 8-bits and is shown in Figure 5. The output of these counters are fed as address input for ROM since the ROM expect the 8-bit address. The ROM address (addr1 [7:0]) increments at every rising edge of clock cycle. However, the other ROM address (addr2 [7:0]) increments while the addr1 reaches its maximum value. This approach produces the matrix 'x' and 'y' as described earlier. The architecture of the dual port ROM is shown in Figure 6 and its signal description is provided in Table 1. The output of ROM generates matrix 'x' and 'y' in a raster scan order. 
      
\begin{figure}
\centering
{
\includegraphics[height=1.5in,width=3in]{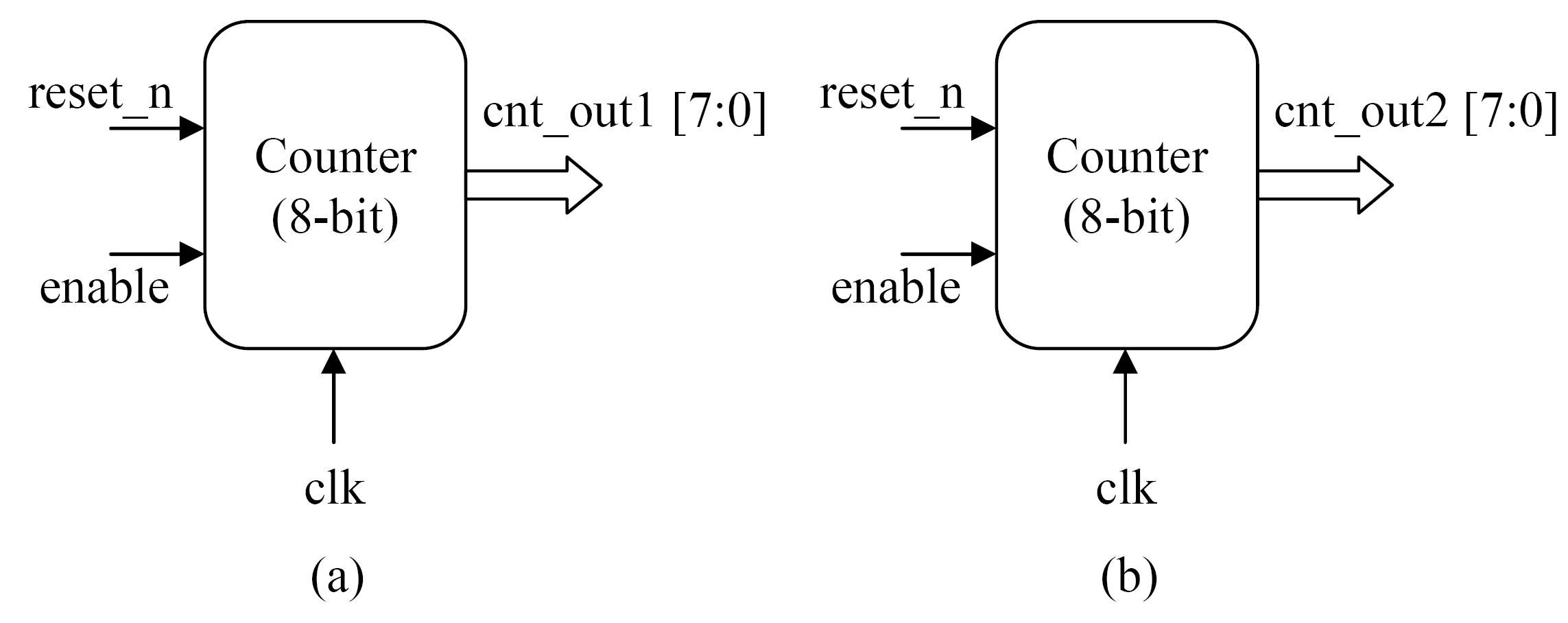}
\label{fig:subfig3}
}
\caption{Design of 8-bit Counter with its Output as Address for Dual Port ROM}
\end{figure}

\begin{figure}
\centering
{
\includegraphics[scale=0.15]{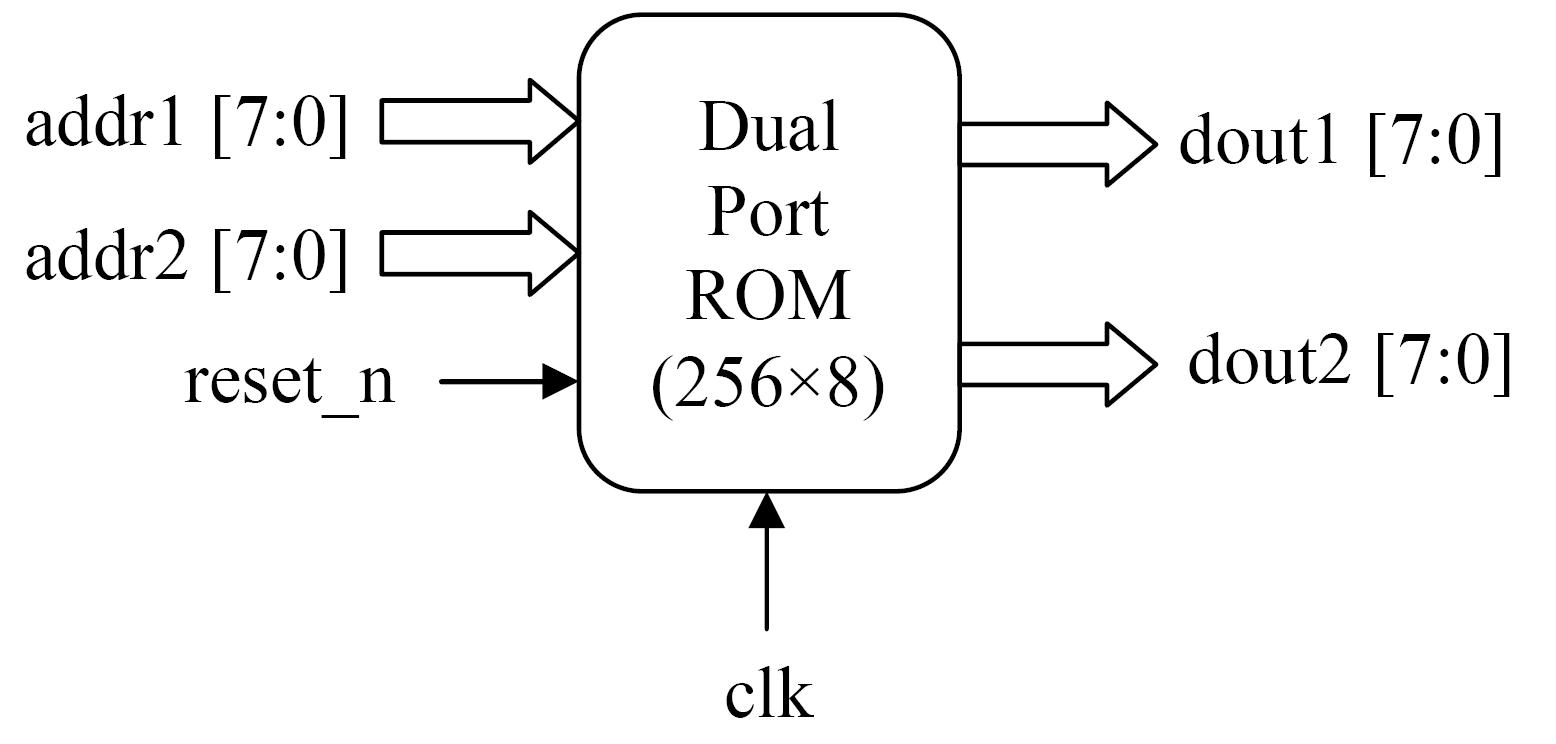}
\label{fig:subfig3}
}
\caption{Architecture of Dual Port ROM}
\end{figure}

% Top Architecture of Multiplier
\begin{figure}
\centering
{
\includegraphics[scale=0.15]{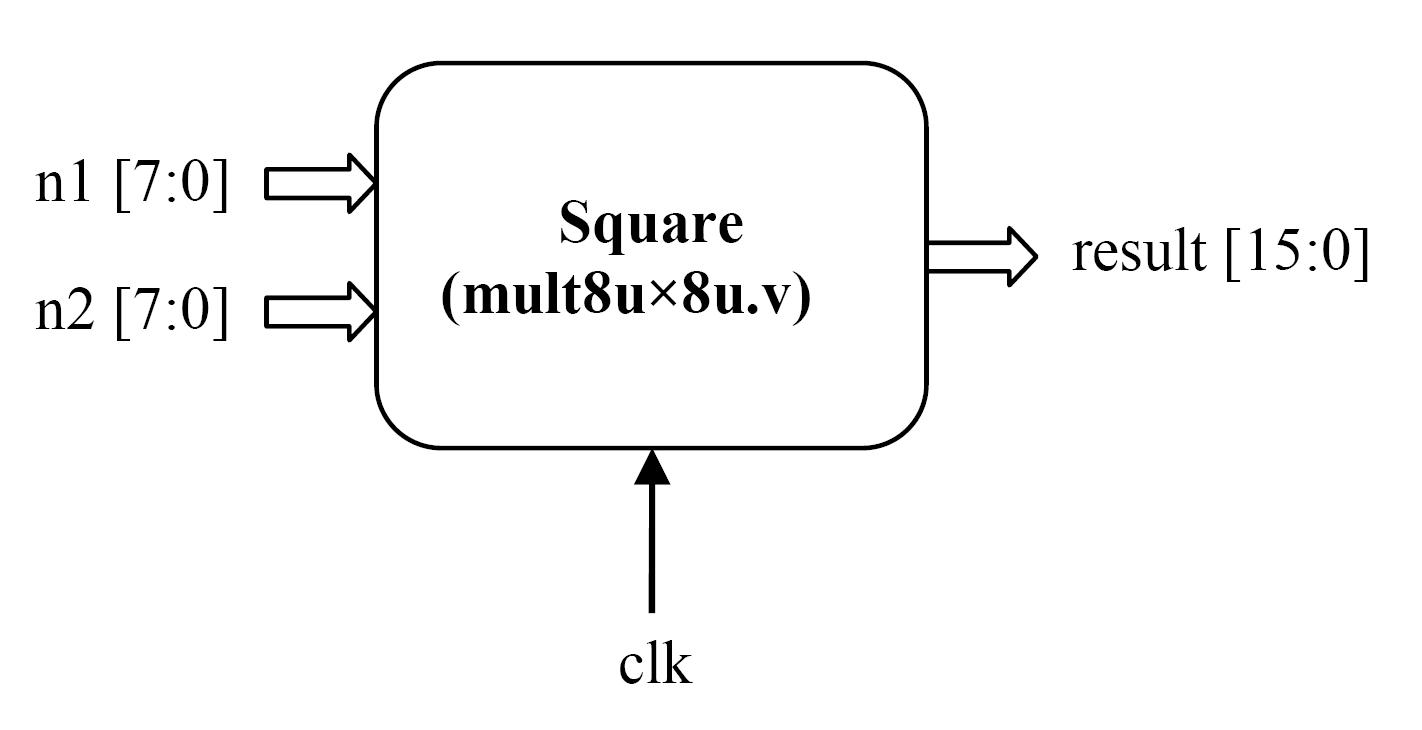}
\label{fig:subfig3}
}
\caption{Top Architecture of $8\times8$ Multiplier}
\end{figure}

The multiplier design \cite{ramachandran} presented in this work incorporates a high degree of parallel circuits and pipelining of five levels. The multiplier performs the multiplication of two 8-bits unsigned numbers n1 and n2 as shown Figure 7 with its signal description in Table 1. The multiplier of width $16\times16$ may also employed for Gaussian function of larger size. The multiplier result is of width 16-bits. The detailed architecture for the multiplier is shown in Figure 8. The architecture utilizes many pipelined registers internally. Five pipelined stages are exploited in order to increase the processing speed.   

% Detailed Architecture of Multiplier
\begin{figure}
\centering
{
\includegraphics[scale=0.1]{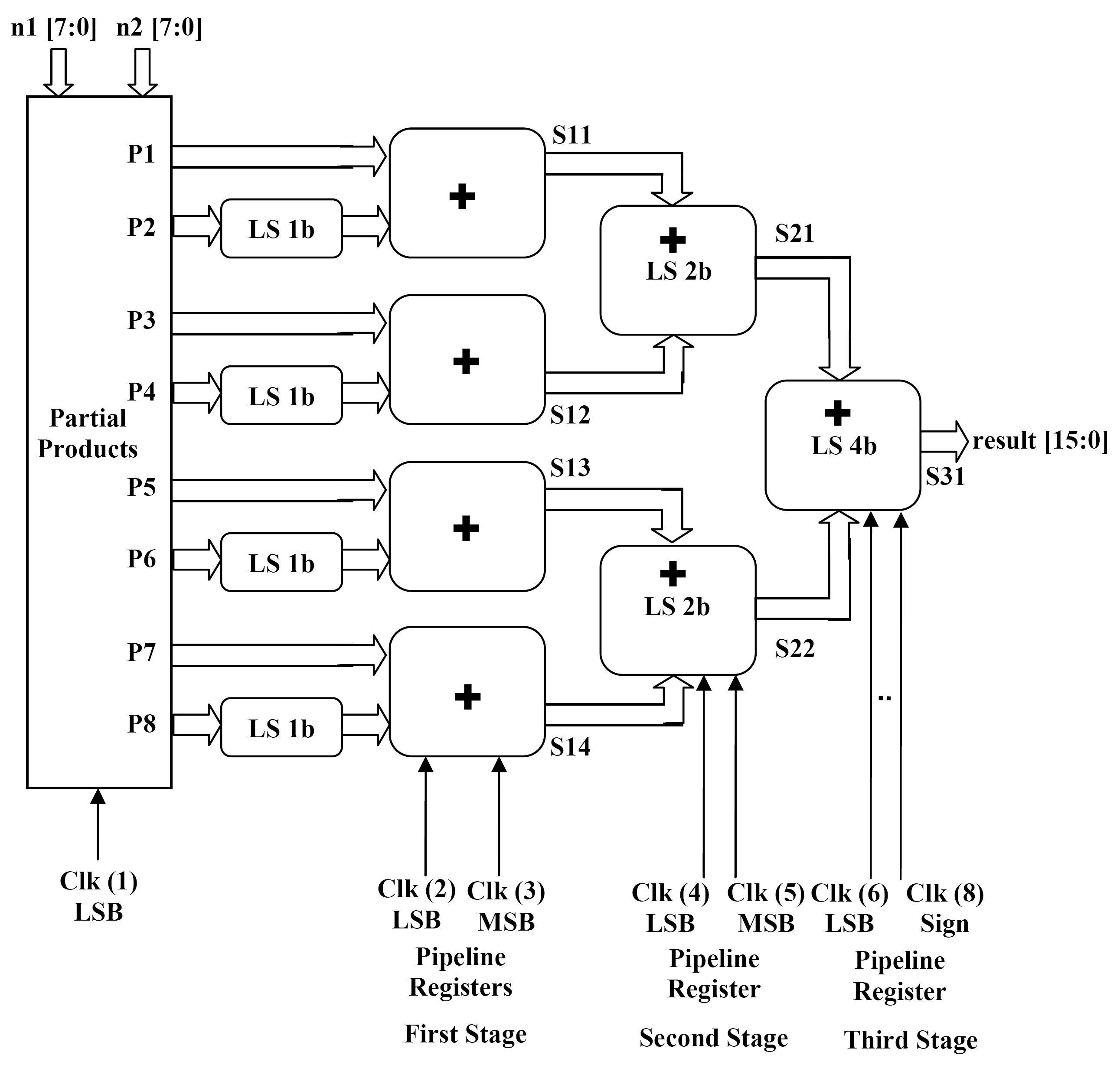}
\label{fig:subfig3}
}
\caption{Detailed Architecture of Multiplier Design with Eight Pipeline Stages}
\end{figure}

\section{Experimental Results and Discussions}

The proposed FPGA implementation of Gaussian surround function has been coded and tested in Matlab (Version 8.1) first in order to ensure the correct working of the algorithm. Subsequently, the complete system has been coded in Verilog HDL so that it may be implemented on an FPGA or ASIC. The proposed scheme has been coded in RTL compliant Verilog and the hardware simulation results have been obtained. The system simulation has been done using ModelSim (Version SE 6.4) and Synthesized using Xilinx ISE 9.1i. The algorithm has been implemented on Xilinx Virtex-II XC2VP40-7FG676 FPGA device. In the proposed work, Gaussian design developed is of size $256\times256$ and further can be upgraded to any size without appreciable increase in the hardware. The functional modules comprising of Gaussian control, ROM, Multiplier, Adder and exponential of Gaussian surround function simulated using ModelSim is presented in Figure 9.  

\begin{figure*}
\centering
\subfigure[Start of Simulation for 2D Gaussian Surround Function]{
\includegraphics[height=2in,width=2.8in]{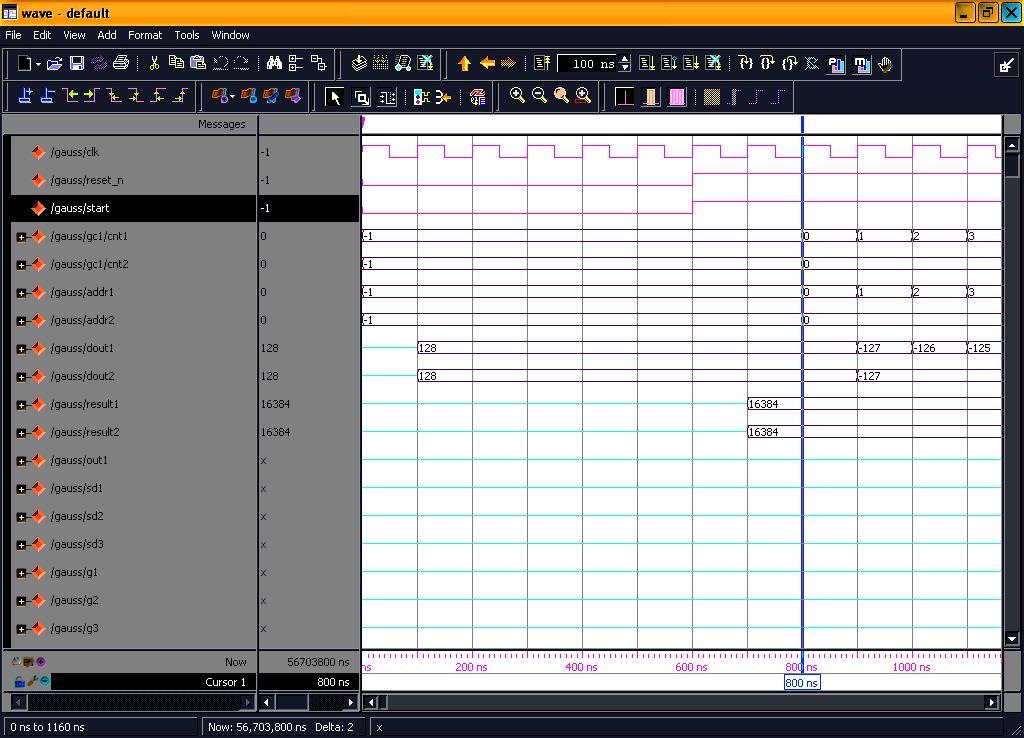}
}
\subfigure[Waveforms for Multiplier Output]{
\includegraphics[height=2in,width=2.8in]{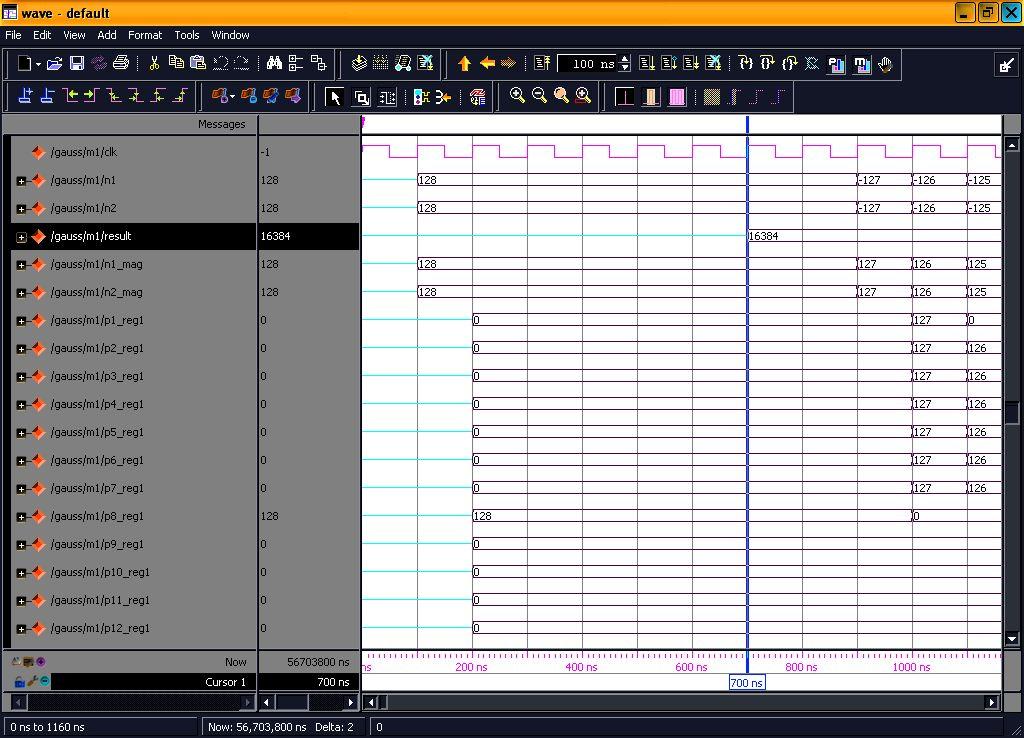}
}
\subfigure[Waveforms for Adder Output]{
\includegraphics[height=2in,width=2.8in]{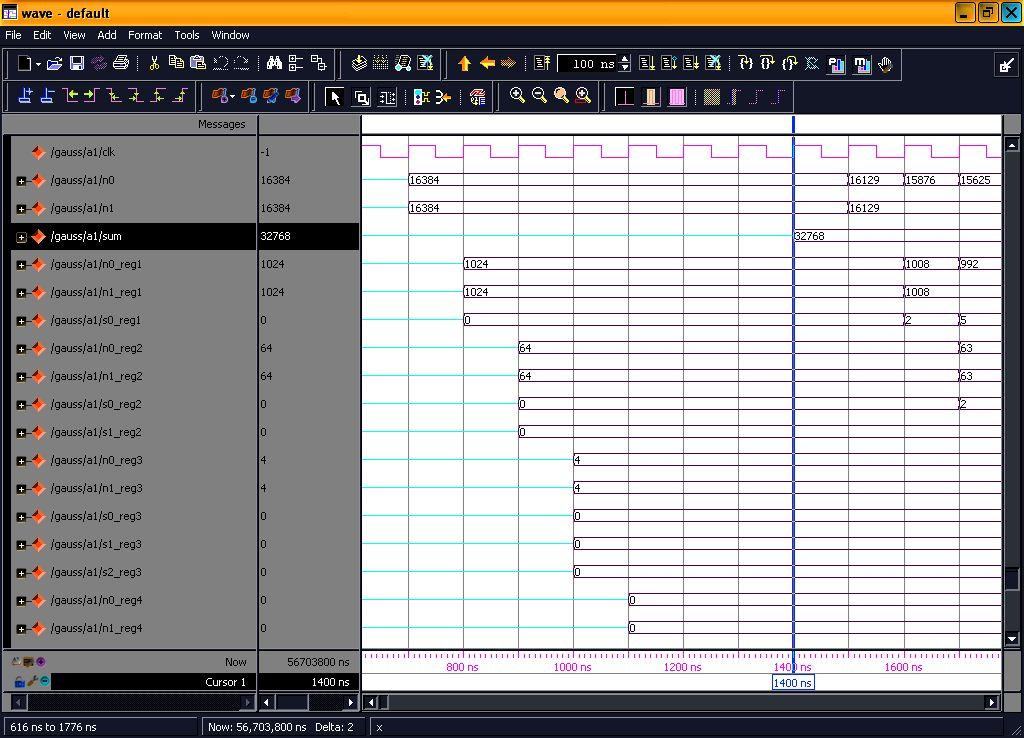}
}
\subfigure[Waveforms for Gaussian Surround Function Output with Three Scales]{
\includegraphics[height=2in,width=2.8in]{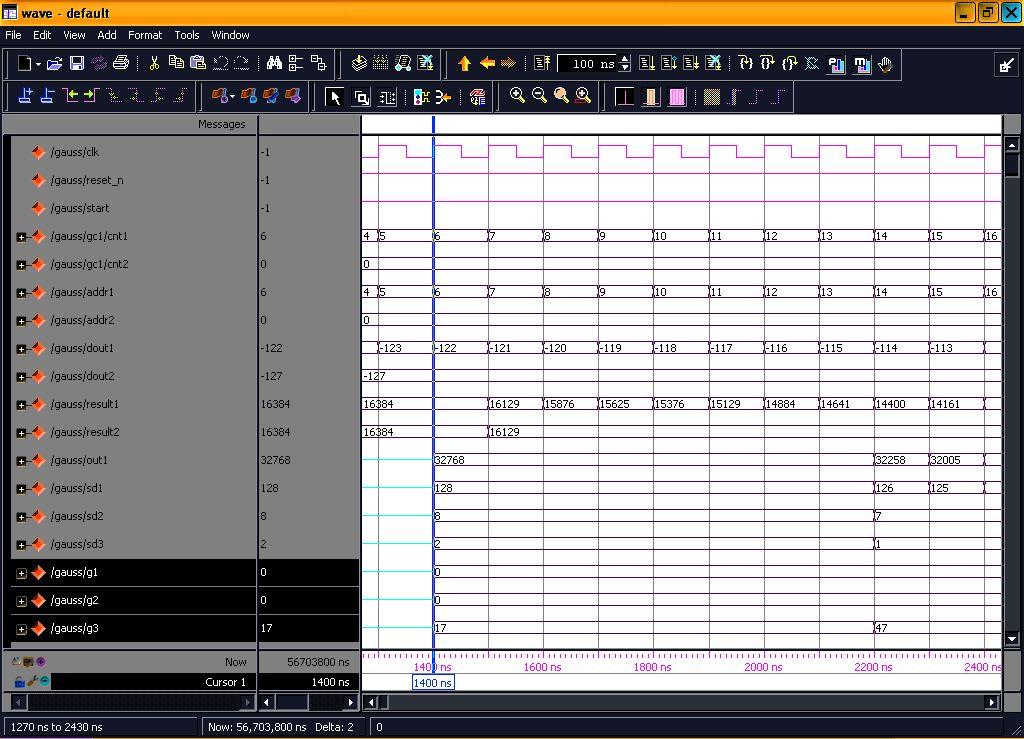}
}
\caption{ModelSim Simulation Waveforms for 2D Gaussian Surround Function}
\label{fig}
\end{figure*}

The Xilinx generated RTL schematic view of top module of "gauss" is shown in Figure 10. The output of this schematic consists of g1 [7:0], g2 [7:0] and g3 [7:0] which is the Gaussian output with three scales namely, 16, 64 and 128. The detailed schematic produced by this top schematic is presented in Figure 11. There are seven modules of which multiplier, and exponent are replicated. The input to the exponent module is in the range of 0 to 4. Therefore, the exponent module is designed with the look-up table technique instead of designing it.  
% RTL Schematics
% Top RTL
\begin{figure*}
\centering
{
\includegraphics[height=1.8in,width=2.3in]{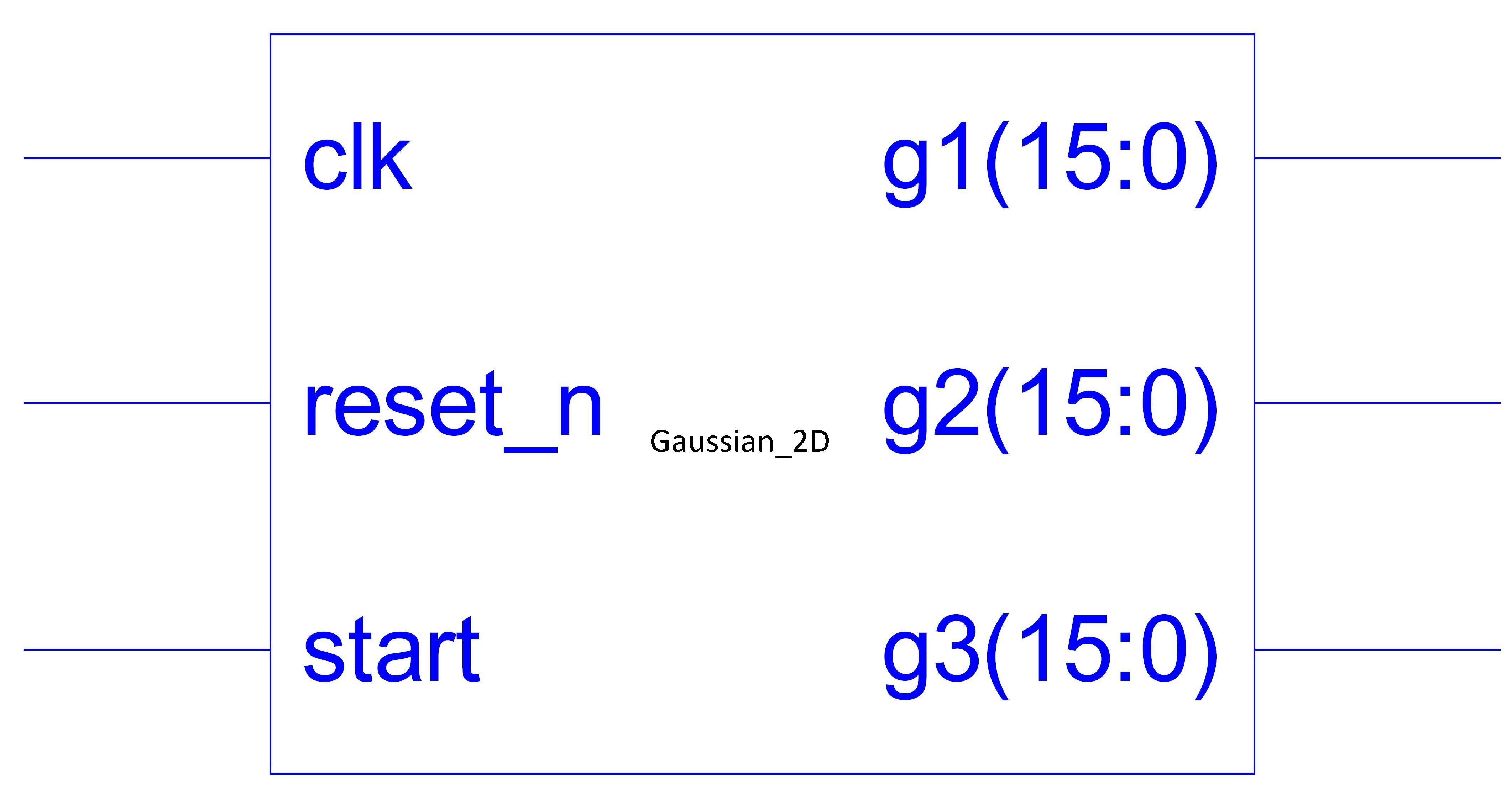}
\label{fig:subfig3}
}
\caption{RTL View of the Top Module "gauss"}
\end{figure*}

% Detailed RTL Schematic
\begin{figure*}
\centering
{
\includegraphics[height=1.7in, width=6in]{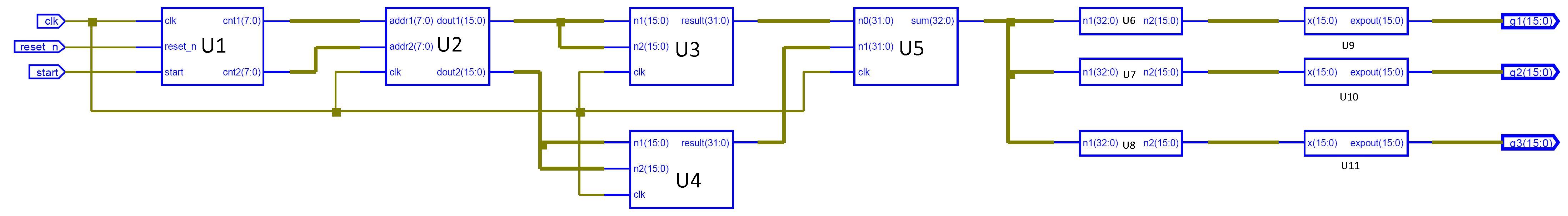}
\label{fig:subfig3}
}
\caption{Zoomed View of the Top Module "gauss"  \textbf{Note:} The module names are not readable in the zoomed view of the Xilinx RTL view. Therefore, in the generated Figure the module names are marked as follows: U1: Gaussian control, U2: Gaussian Memory (ROM), U3 and U4: Multiplier, U5: Adder, U6, U7 and U8: Scale Down Unit, U9, U10 and U11: Exponent}
\end{figure*}

The synthesis as well as place and route results of the Gaussian function is presented in Table 3. The Xilinx place route report for the entire Gaussian function comprising counter, ROM, Multiplier, adder and exponent is shown in Figure 12.

\begin{table*}
\renewcommand{\arraystretch}{1.3}
\caption{Device Utilization Summary}
\label{table47}
\centering
\setlength\tabcolsep{3pt}
  \small
  \begin{tabular}{|l|l|l|}
\hline 
\multicolumn{3}{ |c| }{\textbf{Selected Device : XC2VP40-7FG676}} \\
\hline 
Number of Slices & 725 out of 19,392 & 3\%  \\
\hline 
Number of Slice Flip Flops & 1224 out of 38,784 & 3\%  \\
\hline 
Number of 4 input LUTs & 825 out of 38,784 & 2\%  \\
\hline 
Number used as logic & 713 &   \\
\hline 
Number used as Shift registers & 94 &  \\
\hline 
Number of IOs & 51& \\
\hline 
Number of bonded IOBs &51 out of 416 & 12\% \\
\hline 
Number of GCLKs & 1 out of 16 & 6\%  \\
\hline 
\end{tabular}
\end{table*}

The timing summary for the design as reported by Xilinx ISE tool is as follows:
%\\Speed Grade : -7
%\\Minimum period: 4.483 ns (Maximum Frequency: 223.04 MHz)
%\\Minimum input arrival time before clock:     1.473 ns
%\\Maximum output required time after clock:  6.880 ns
%\\Timing constraint: Default period analysis for Clock 'clk'
%\\Clock period: 4.483 ns (frequency: 223.04 MHz)
%\\Total number of paths /destination ports: 26126 / 1922
%\\Delay: 4.483 ns (Levels of Logic = 18)

\begin{Verbatim}[frame=single]
1. Speed Grade : -7
2. Minimum period: 4.483 ns 
(Maximum Frequency: 223.04 MHz)
3. Minimum input arrival time 
before clock: 1.473 ns
4. Maximum output required time 
after clock:  6.880 ns
5. Clock period: 4.483 ns 
(frequency: 223.04 MHz)
6. Total number of paths /destination 
ports: 26126 / 1922
7. Delay: 4.483 ns 
(Levels of Logic = 18)
\end{Verbatim}

% FPGA Implementation

\begin{figure*}
\centering
\subfigure{
\includegraphics[height=0.7in,width=4.5in]{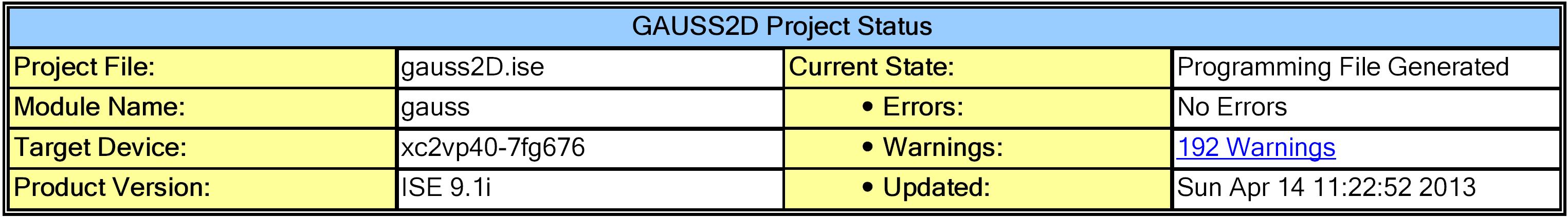}
}
\subfigure{
\includegraphics[height=2in,width=4.5in]{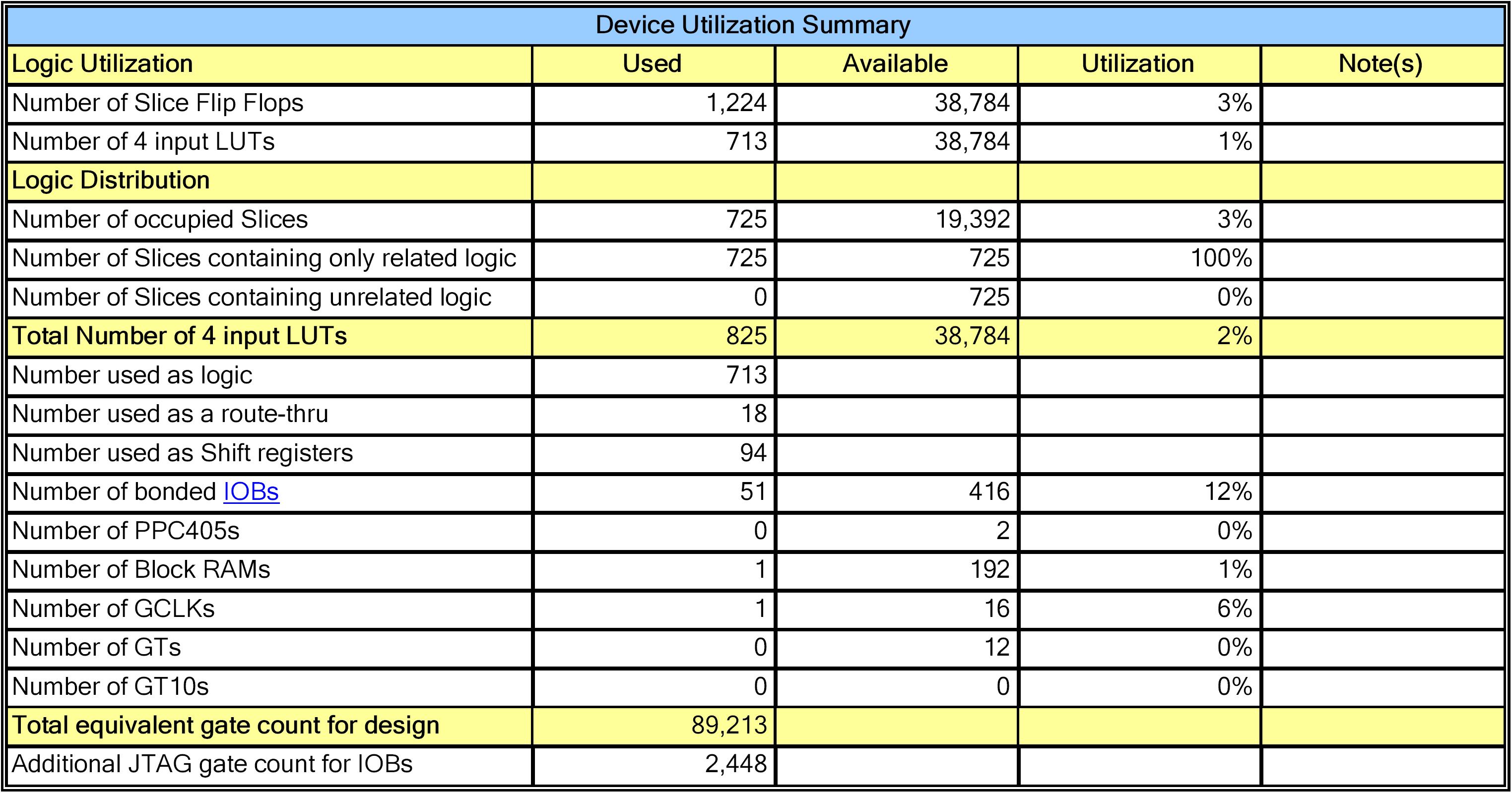}
}
\caption{Device Utilization Summary as Reported by Xilinx Tool}
\label{fig}
\end{figure*}

\begin{table*}
\renewcommand{\arraystretch}{1.3}
\caption{Design Mapped on Various FPGA Devices}
\label{table47}
\centering
\setlength\tabcolsep{2pt}
  \begin{tabular}{|l|l|l|l|l|}
\hline 
Device & LUTs & Gates & Utilization& Frequency \\
\hline
XC2S50-6TQ144 & 825 out of 1536 & 62,455& 53\% &61.992 MHz \\
\hline 
XC2S200-6PQ208 & 825 out of 4704 & 62,455& 17\% &61.992 MHz \\
\hline 
XC3S1600E-5FG484 & 825 out of 29504 & 89,213& 5\% &160.458 MHz \\
\hline 
XC2VP40-7FG676 & 825 out of 38784 & 89,213& 2\% &223.04 MHz \\
\hline 
\end{tabular}
\end{table*}

\section{Conclusion}
The 2D Gaussian surround function has been designed for processing high resolution pictures of size $1600\times1200$ at real time rate of 30 frames per second. The Gaussian function design developed is based on the symmetry property which consumes low on-chip memory. The 2D Gaussian function implementation is presented for use in applications such as image
enhancement, smoothing, edge detection and filtering etc. The architectures developed was coded in Verilog, conforming to RTL coding guidelines used in industries and fits onto a single chip with a gate count utilization of 89,213 gates. Research work is in progress for employing the Gaussian surround function for Multiscale Retinex based color image enhancement.

\pagebreak
\vspace{6cm}
\noindent{\includegraphics[width=1in,height=1.7in,clip,keepaspectratio]{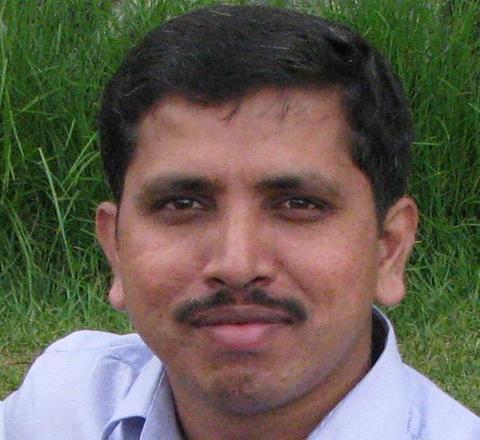}}
\begin{minipage}[b][1in][c]{1.8in}
{\centering{\bf {M. C Hanumantharaju}} is currently a Associate Professor in the Department of Information Science \& Engineering, Dayananda Sagar College of Engineering, Bangalore. He is currently persuing Ph.D at Visvesvaraya Technological University,Bel-}
\end{minipage}\\\\ gaum His research interests includes VLSI Architecture Development for Signal \& Image Processing Applications, Synthesis \& Optimization of ICs, DSP with FPGAs etc.
 \\\\\\

%---------------------------------------------------------------------------------------------------------------------
%\begin{biography}
\noindent{\includegraphics[width=1in,height=1.8in,clip,keepaspectratio]{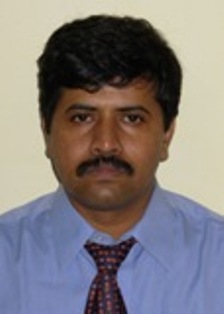}}
\begin{minipage}[b][1in][c]{1.9in}
{\centering{\bf{M. T Gopalakrishna }}is currently a Associate Professor in the Department of Information Science \& Engineering, Dayananda Sagar College of Engineering, Bangalore. He is currently persuing his Ph.D at Visvesvaraya Technological University, Belgaum. His Research interests includes Digital Image Processing \& Computer Vision, Video Surveillance and Document Image Processing.}
\end{minipage}
\\\\

\end{document}